\documentclass[aps,twocolumn,showpacs,superscriptaddress]{revtex4}

\usepackage{graphicx}%
\usepackage{dcolumn}
\usepackage{amsmath}

\begin{document}


\title{2D Metal-Insulator transition as a percolation transition}

\author{S. Das Sarma}
\affiliation{Condensed Matter Theory Center, Department of Physics,
University of Maryland, College Park, MD 20742}
\affiliation{Sandia National Laboratories, Albuquerque, NM 87185}%

\author{M. P. Lilly}
\affiliation{Sandia National Laboratories, Albuquerque, NM 87185}%

\author{E. H. Hwang}
\affiliation{Condensed Matter Theory Center, Department of Physics,
University of Maryland, College Park, MD 20742}

\author{L. N. Pfeiffer}
\author{K. W. West}
\affiliation{Bell Laboratories, Lucent Technologies, Murray Hill, NJ 07974}

\author{J. L. Reno}
\affiliation{Sandia National Laboratories, Albuquerque, NM 87185}%

\date{\today}

\begin{abstract}
By carefully analyzing the low temperature density dependence of 2D
conductivity in undoped high mobility n-GaAs heterostructures, we
conclude that the 2D metal-insulator transition in this system is a density
inhomogeneity driven percolation transition due to the breakdown of
screening in the random charged impurity disorder background. In
particular, our measured conductivity exponent of $\sim 1.4$ approaches
the 2D percolation exponent value of 4/3 at low temperatures and our
experimental data are inconsistent with there being a zero-temperature
quantum critical point in our system.
\end{abstract}

\pacs{71.30.+h, 73.40.-c, 73.50.Bk}

\maketitle

Ever since the pioneering observation by Kravchenko {\it et al}.
\cite{kravchenko1} of an apparent two-dimensional (2D) metal-insulator
transition (MIT) in Si-MOSFET inversion layers, the nature of the
transition has remained a controversial enigma. The 2D MIT has been
claimed by some, most notably by Kravchenko and collaborators
\cite{kravchenko1,abrahams}, to be an interaction driven and carrier density tuned
T=0 quantum phase transition whereas others have argued \cite{altshuler}
that it is a disorder-driven crossover phenomenon between weakly and
strongly localized 2D electron states. Density dependent scanning
studies of the 2D chemical potential \cite{yacoby}, surface acoustic
wave measurements on low density 2D electrons\cite{eisenstein_saw}, and
compressibility measurements \cite{eisenstein,jiang}, as analyzed by
direct numerical simulations \cite{shi} of the energetics of the 2D
disordered system, indicate very strong density inhomogeneities in the
2D system around the critical density ($n_c$) for the putative 2D MIT.
Also, the critical density $n_c$ is highly system-dependent and varies
very strongly with the impurity disorder in the 2D system. In relatively
highly disordered Si MOSFETs the typical $n_c \approx 10^{11} \mbox{cm}^{-2}$
and in very high quality 2D n-GaAs heterostructures (the subject matter
of our work presented in this paper) the typical $n_c \sim 10^9
\mbox{cm}^{-2}$. The strong dependence of $n_c$ on disorder, the long range
nature of the charged impurity disorder potential in 2D semiconductor
structures, the observed density inhomogeneities around $n_c$, all taken
together, suggest a percolation-type transition underlying the 2D MIT.
Indeed such a percolation transition for 2D systems was discussed by
Efros fifteen years ago \cite{efros1}, and has
recently been suggested in the context of the 2D MIT phenomena by
several authors \cite{shi,he,meir,fogler,dassarma1}. In this Letter we
provide rather compelling experimental evidence in support of the 2D MIT
being a percolation transition, at least for our experimental system.

The basic picture of the percolation transition for 2D MIT is simple and
highly physically motivated. As the carrier density ($n$) is lowered in
a 2D system, screening becomes progressively weaker and strongly
nonlinear. A small decrease in $n$ leads to a large decrease in
screening, and eventually to a highly inhomogeneous 2D system as the
electron gas at low enough carrier densities becomes unable to screen
the disorder potential with the individual charged impurity scattering
potentials getting unscreened and their random distribution in the 2D
system.  This gives rise to the well-known random ``hill-and-valley''
potential landscape with the 2D carriers repelled from the potential
hills and accumulating at the potential valleys in contrast to the high
carrier density homogeneous situation. Once these ``depletion (or
denuded) zones'', associated with the disorder potential hills are
numerous enough to prevent percolating conducting paths to span the 2D
system connecting the ``valley'' regions, an effective 2D MIT
transition takes place with the system being an effective
metal (insulator) for $n > (<)$ $n_c$ where $n_c$ is the critical
percolation density. This percolation picture is particularly germane to 2D
semiconductor systems because the disorder potential here arises from
the presence of random charged impurities in the system making
electronic screening a key ingredient in the effective disorder seen by
the carrier system. Such a density-driven percolation transition is the
hallmark of the long-range Coulomb disorder potential in the system
since the Coulomb disorder can be tuned by changing the carrier density
through the nonlinear screening mechanism. The key qualitative features
of the 2D MIT phenomenon, namely the correlation between $n_c$ and the
sample quality as well as the density inhomogeneity around $n_c$, are
entirely consistent with a percolation transition in the screened
Coulomb disorder potential. 

Motivated by these percolation considerations we have carefully
analyzed the density dependent conductivity $\sigma(n)$
in ultra-high mobility n-GaAs undoped heterostructures which have demonstrated
\cite{lilly} the lowest value of $n_c$ ($\sim 2 \times 10^9 \mbox{cm}^{-2}$) 
for 2D MIT
so far in the literature. Most of the 2D MIT literature has emphasized
the strong temperature dependence in the effective metallic ($n>n_c$)
phase, which is now well-understood \cite{dassarma2} as a manifestation of
the strong temperature dependence of the effective Coulombic disorder
seen by the low-density 2D carriers through the temperature dependence
of screening. In this Letter, we analyze the power law behavior of
$\sigma(n)$ in the context of density dependent screening and
a percolation transition. We have
also carried out a detailed Boltzmann theory calculation \cite{dassarma2}
using realistic quasi-2D parameters appropriate for our system,
assuming the carrier transport to be dominated by charged impurities
distributed randomly at the interface -- the main physics being the
temperature and wave vector dependent linear screening of the long-range
Coulomb scattering potential of the charged impurities. We have
earlier shown \cite{lilly} that such a screening description provides an
excellent qualitative agreement with experiment for the temperature
dependent transport properties of the effective metallic phase in our
system.

The two samples used in this study are undoped GaAs/AlGaAs
heterojunctions with field-induced carriers. From the surface, the
epilayers consist of 60 nm n+ GaAs followed by a 600 nm
Al$_{0.3}$Ga$_{0.7}$As barrier, and then 1000 nm of GaAs. The 2D
electrons are induced from the ohmic contacts with a density that is
proportional to the voltage on the n+ GaAs layer that serves as a top gate.
Standard four terminal lock-in measurements are made to measure the
resistance over a wide range of density. Conductivity of these square
samples is determined using van der Pauw techniques. In Sample A the
density spans a range of $0.15$ to $7.5 \times 10^{10} \mbox{cm}^{-2}$
with a very high maximum mobility of $8.5 \times 10^{6}
\mbox{cm}^2/\mbox{Vs}$ at high density and sample B spans a density
range of $0.5$ to $13 \times 10^{10} \mbox{cm}^{-2}$ with a lower
mobility of $3.3 \times 10^{6} \mbox{cm}^2/\mbox{Vs}$ at high
density. 

\begin{figure}
\includegraphics{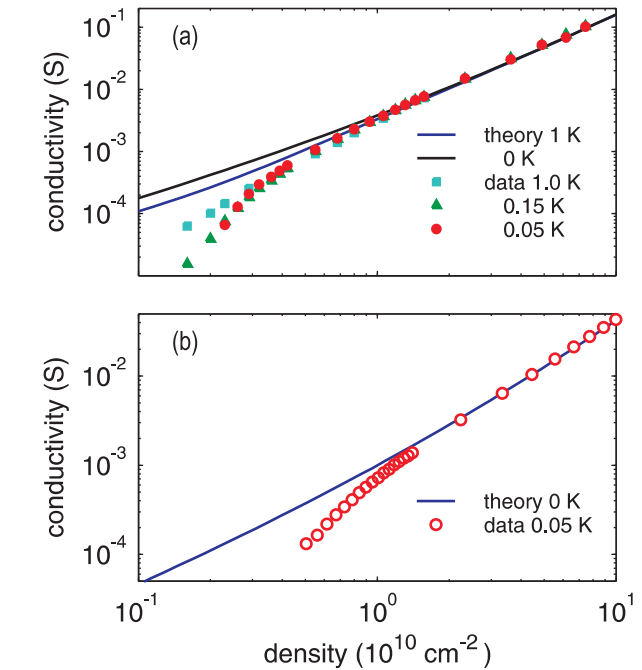}
\caption[fig1]{Experimentally measured (symbols) and Boltzman scattering calculated 
(lines) values of conductivity for sample A (1a) and sample B (1b).}
\end{figure}

The experimentally measured conductivity $\sigma(n)$ is shown in Fig. 1
along with the calculated\cite{dassarma2} $\sigma(n)$ curves. We first
discuss a few salient qualitative features of the density dependent
transport properties of our results. At high
densities ($n \stackrel{>}{_\sim} 10^{10} \mbox{cm}^{-2}$), the
conductivity depends on the density approximately as $\sigma \sim
n^{\alpha}$ with $\alpha \approx 1.6$. We emphasize that this is not
a strict power law since $\alpha \equiv \alpha(n)$ depends weakly on the
density. This high density behavior is completely consistent with our
Boltzmann theory based calculations assuming the conductivity being
limited by linearly screened charged impurity scattering. As $n$
decreases, $\sigma$ starts decreasing faster with decreasing density and
the experimental conductivity exponent $\alpha(n)$ becomes strongly
density dependent with its value increasing substantially above the
high-density value of $\alpha \approx 1.6$. At the lowest density for
sample A, $\alpha \approx 5$. Although a part of this strong density
dependence of $\alpha(n)$ at low density can be understood as arising
from the strong suppression in screening at low carrier densities, we
find that screening in a $homogeneous$ electron gas fails
qualitatively in explaining the $\sigma(n)$ behavior at low densities
whereas it gives quantitatively accurate results at high densities. As
has been found from direct numerical simulations \cite{shi,efros1},
homogeneous screening of charged impurity disorder breaks down at low
carrier densities with the 2D system developing strong inhomogeneities
leading to a percolation transition at $n=n_c$. For $n <n_c$, the system
is an insulator containing isolated puddles of electrons with 
no ``metallic'' conducting path spanning through the
whole system.

The percolation scenario also naturally explains the qualitative 
aspects of the observed strong temperature dependence of the effective 
metallic phase in low-disorder high-quality samples for 
$n \stackrel{>}{_\sim} n_c$ in the 
2D MIT phenomenon.  In particular, low disorder ensures a low value 
of $n_c$, which then 
automatically leads to a strongly temperature dependent screening 
correction producing a temperature dependent conductivity as both $T/T_F$ 
and $q_{TF}/2k_F$ are effectively large for low densities\cite{dassarma2}.  
In samples with large 
values of $n_c$ (low-quality samples) the temperature dependent screening 
effects are weak in the effective metallic phase ($n>n_c$) since the 
carrier densities are high with the associated $q_{TF}/2k_F$ and 
$T/T_F$ being ``small''\cite{dassarma2}.

\begin{figure}
\includegraphics{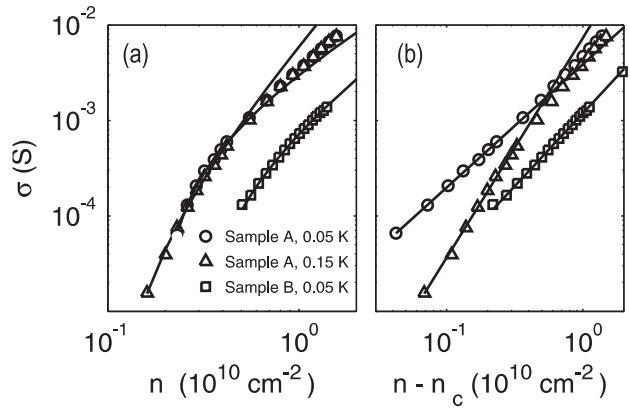}
\caption[fig2]{Experimental conductivity (symbols) fit to Eq. 1 (lines)
for both Sample A and B as indicated in the legend. The same results are
displayed as a function of $n$ (2a) and $n - n_c$ (2b) to better show
the low density region.}
\end{figure}

In Fig. 2 we fit our measured conductivity $\sigma(n)$ to the
expected\cite{efros1,stauffer} percolation `critical'
behavior \begin{equation} \sigma(n) = A (n-n_c)^{\delta},
\end{equation} where $\delta$ is the conductivity percolation exponent
characterizing the vanishing of the conductivity. Nonlinear
curve fitting to Eq. (1) with proper weighting of the conductivity is
performed for a range of the low density data to determine values
of $\delta$, $n_c$ and $A$. For sample A the data used in the fit
is $1 \times 10^9 < n < 4 \times 10^9 \mbox{cm}^{-2}$,
although the results do not depend sensitively on the exact
range of data used in the fit. Note that for T = 0.05 K and $n < 2.3
\times 10^9 \mbox{cm}^{-2}$ a large lock-in quadrature signal prevents
measurement of the conductivity. For Sample B, the `goodness' of the fit
is more strongly dependent on the range of data used for fitting. For the
data shown, the lowest density point is excluded from the fit, and the
fitting range is $5 \times 10^9 < n < 8 \times 10^9 \mbox{cm}^{-2}$. The
values of the relevant parameters are $\delta = 1.4 \pm 0.1$ and
$n_c = 0.18 \pm 0.01 \times 10^{10} \mbox{cm}^{-2}$ at $T=50$ mK for
sample A and $1.5 \pm 0.1$ and $n_c = 0.28 \pm 0.02 \times 10^{10} \mbox{cm}^{-2}$
for sample B. These values of $\delta$ are close to the
known 2D percolation exponent of 4/3\cite{stauffer}. 

In Fig. 3 we show the dependence of the critical exponent $\delta(T)$
and the critical density $n_c(T)$ of sample A. At low
temperature, the experimental conductivity exponent $\delta(T)$
approaches a value $\delta \approx 1.4$ (1.5 for sample B), which is
close to the expected 2D percolation exponent. 
The decrease of $\delta$ with T for $T >0.2
K$ is caused by the increasing importance of phonon scattering which
should be disregarded for our discussion of the 2D MIT itself. The
increase of $\delta$ (for $T<0.2$ K) and decrease of $n_c$ with
increasing temperature can be well-understood by adding a non-critical
temperature dependent contribution $f(T)$, where $f(T \rightarrow 0) =0$
and $f(T) \neq 0$ for $T \neq 0$, to the right-hand side of Eq. 1 and
realizing that at any finite T the system, by definition, is an
``effective'' metal since the conductivity is non-zero at all finite
temperatures. This automatically implies that the effective $\delta(T)$
increases with T and the effective $n_c(T)$ decreases with T, since
$\delta$ and $n_c$ are strictly defined only at $T=0$ and at all finite
T the system tends to behave as an effective metallic phase -- in fact,
for sufficiently high values of T ($\ge 1K$) one cannot distinguish at
all between the metallic and the insulating phase by looking at the
conductivity. The fact that critical behavior manifests over about a
decade of density and two decades of conductivity as well as our finding
of a conductivity exponent consistent with the 2D percolation transition
suggests that the 2D MIT in these low disorder n-GaAs systems is a percolation 
transition. 

\begin{figure}
\includegraphics{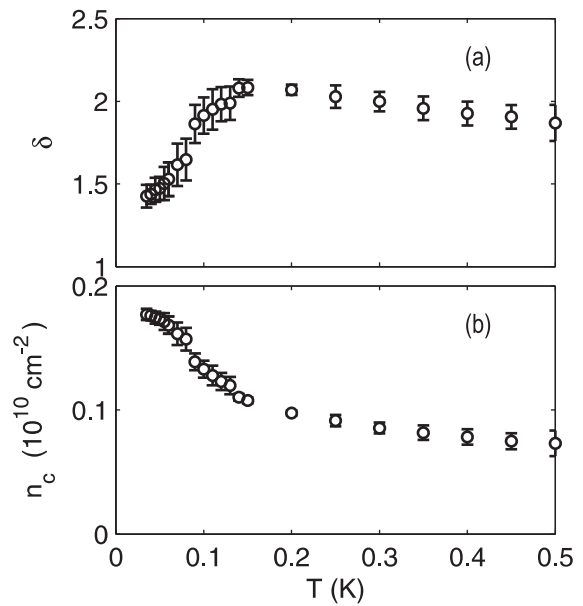}
\caption[fig3]{Parameters $\delta$ and $n_c$ for sample A for a range of
temperature. Error bars are the 95\% confidence interval as determined
from the nonlinear fitting shown in Fig. 2. Sample B shows much weaker
temperature dependence of $\delta$ and $n_c$ in this range.}
\end{figure}

It is important to point out that the critical density $n_c$ we obtain
by the optimal fitting of experimentally measured $\sigma(n)$ to the
conductivity scaling formula of Eq. 1 is consistently lower than the
corresponding estimate ($n_c'$) based on the sign of $d\sigma/dT$ as is
often done in the literature to distinguish an effective metal from the
insulator. This is perfectly understandable based on the
``quantum-classical crossover'' scenario discussed in ref.
\onlinecite{dassarma2} -- at low carrier densities the effective
metallic phase exhibits an insulating $d\sigma/dT$ ($>0$) sign down to
rather low temperature, and therefore any determination of the critical
density $n_c'$ based purely on the sign of $d\sigma/dT$ will produce
$n_c' >n_c$. Thus, the 2D MIT transition density cannot be ascertained
by looking for the point $d\sigma/dT|_{n=n_c'} = 0$ -- the sign of
$d\sigma/dT$ is {\it not} a good indicator for the critical density. Our
analyses as shown in Figs. 1-3 imply that the transition itself is
better understood by concentrating on the density dependence of
$\sigma(n)$ at fixed (and very low) values of temperature whence the 2D
MIT can be studied approaching entirely from the metallic regime with
the sign of $d\sigma/dT$ playing no role in the analysis. The effect of
electron-electron interaction (beyond just the nonlinear screening
mechanism leading to the electron 'puddles' or 'droplets') is mainly to
decrease the effective value of the critical density at which the
percolation transition occurs, since the main effect of interaction is
to 'homogenize' the density of the electron liquid in the process
reducing the effect of density inhomogeneities arising from the random
charged impurity centers. Without interaction effects, percolation will
occur at higher carrier densities\cite{shi}.

Before concluding we emphasize that one obvious and immediate
consequence of the 2D MIT being a disorder driven percolation transition in our system
is that it is manifestly {\it not} a quantum critical phenomenon as has
sometimes been suggested in the literature. We have, in fact, attempted
to fit our density and temperature dependent conductivity $\sigma(n,T)$
to the usual quantum critical scaling form \cite{kravchenko1,abrahams},
$\sigma(n,T) \sim F(T/T_0)$, where $F$ is the universal scaling function
with $T_0 \equiv T_0(n) \sim |n-n_c|^{\gamma}$ where the exponent
$\gamma \equiv \nu z$ is a product of the correlation exponent $\nu$ and
the dynamical exponent $z$. We have not been able to obtain any kind of
scaling collapse of our $\sigma(n,T)$ data as a function of $T/T_0$. The
inconsistency of our results with quantum criticality is obvious from
the percolation scaling fits to Eq. (1) that our data exhibit so well.
The fact that our fitted percolation exponent $\delta$ and critical
density $n_c$ both depend on temperature already indicates an absence of
quantum critical scaling in the data. 
A more subtle point is that the
usual quantum critical `fan' diagram\cite{sondhi} indicates that the scaling regime
should expand with increasing temperature (at low enough
temperatures) since the quantum critical regime spreads out in density (around
$n_c$) at higher values of T. The expected quantum critical `fan'
behavior is in strong disagreement with our observed
behavior of $\sigma(n,T)$ where we find that the best scaling
of $\sigma(n) \sim (n-n_c)^{\delta}$ occurs at our lowest measurement
temperatures, and the scaling regime (i.e. the regime in carrier
density) definitively and systematically shrinks as temperature
increases in sharp contrast to the quantum scaling fan behavior. It
cannot, of course, be ruled out that the quantum critical regime ends
at temperatures way below our lowest measurement temperature ($\sim 50
$ mK).   We believe that our
data and the percolation analyses presented in this paper
rules out a quantum critical phenomenon underlying 2D
MIT.

We conclude by emphasizing that we provide in this Letter compelling
evidence involving both experimental transport data and detailed
theoretical analyses in support of the 2D MIT phenomenon being a 2D
percolation transition in our system where the carrier system breaks up into an
inhomogeneous collection of disconnected local `droplets' of electrons
localized in the potential valleys of the long-range Coulombic disorder
potential. Our measured conductivity exponent $\delta \approx 1.4 -
1.5$, approaches the expected 2D percolation exponent of 4/3, and we see
impressive scaling behavior in our data consistent with a percolation
transition. The percolation transition occurs at a critical density
$n_c$ considerably lower than the corresponding `critical' density
$n_c'$ one would infer for 2D MIT based just on the temperature
dependence of $\sigma(n,T)$. By comparing with our theoretical
calculations, we establish that the effective disorder underlying 2D MIT
is the scattering from screened charged impurity centers randomly
distributed near the semiconductor-insulator interface. As screening
weakens at low carrier densities, the system breaks up into
inhomogeneous carrier droplets separated by denuded or depleted
zones, leading eventually to the percolation transition as was discussed
by Efros some years ago \cite{efros1}. Our transport data validates the
existence of a percolation transition in 2D MIT but cannot really
provide a qualitative description of the underlying percolation
process -- for example, we cannot rule out percolation scenarios which
are somewhat different from the physically appealing (but perhaps
somewhat simplistic) picture of Efros where the percolation transition
occurs primarily through the depleted zones becoming numerous enough at
low enough carrier densities. More experimental work, involving
simultaneous measurements of compressibility, capacitance, transport, local chemical
potential, and perhaps even local conductivity, is clearly required to
better understand the details of the percolation mechanism and the
nature of the inhomogeneous 2D carrier system. Our work also brings up important questions
regarding the universal nature of 2D MIT itself -- for example, we urge
experimental analyses of $\sigma(n)$ at fixed low temperatures in {\it other}
2D systems to check whether this percolation scenario is universal or just
applies to our ultra-high mobility 2D n-GaAs system. In this context it is
perhaps quite interesting to point out that even the extensively studied
3D MIT in doped Si systems has been speculated\cite{bogdanovich} to be
a possible inhomogeneity driven percolation transition (similar to what
we find for 2D MIT in our work), and such a percolation transition may
very well be a generic mechanism for metal-insulator transitions in
semiconductors where screened charged impurity potential is the dominant
disorder mechanism. 

We acknowledge outstanding technical assistance from Denise Tibbetts and
Roberto Dunn.  This work (SDS and EH) is supported by the US-ONR, US-ARO, 
and NSF at the University of Maryland.  This work was performed, 
in part, at the Center for Integrated
Nanotechnologies, a U.S. Department of Energy, Office of Basic Energy
Sciences nanoscale science research center operated jointly by Los
Alamos and Sandia National Laboratories. Sandia National Laboratories is
a multi-program laboratory operated by Sandia Corporation, a
Lockheed-Martin Company, for the U. S. Department of Energy under
Contract No. DE-AC04-94AL85000.

\end{document}